\documentclass[floatfix,showpacs,twocolumn,a4paper,pra]{revtex4}
\usepackage{graphicx,amsmath}

\begin{document}

\title{Field correlations and effective two level atom-cavity systems}
\author{S. Rebi\'{c}}
\altaffiliation[Present address: ]{INFM and Dipartimento di Fisica, Universit\`{a} di Camerino, I-62032 Camerino
(MC), Italy}
\email[E-mail: ]{stojan.rebic@unicam.it}
\author{A. S. Parkins}
\author{S. M. Tan}
\affiliation{Department of Physics, University of Auckland,
Private Bag 92019, Auckland, New Zealand}

\date{\today}

\begin{abstract}
We analyse the properties of the second order correlation functions of the
electromagnetic field in atom-cavity systems that approximate two-level
systems. It is shown that
a recently-developed polariton formalism can be used to account for all
the properties of the correlations, if the analysis is extended to include
two manifolds - corresponding to the ground state and the states excited
by a single photon - rather than just two levels.
\end{abstract}
\pacs{42.50.Dv, 32.80.Qk, 42.50.Lc}
\maketitle

The fundamental challenge for nonlinear quantum optics is the realization of dissipation-free photon-photon
interactions at the level of a few photons. In conventional nonlinear optical systems, Kerr nonlinearity gives
rise to an effective photon-photon interaction that becomes important typically on the level of $10^{10}$
photons. Enhancement of the atom-field coupling using the techniques of cavity quantum electrodynamics (cavity
QED) increases  the Kerr nonlinearity. Two spectacular experiments have demonstrated that it is indeed possible
to obtain large {\em conditional phase shifts} that arise from strong photon-photon interactions at the few
photon level~\cite{Kimble95}. The basic cavity QED scheme utilized in these experiments is based on the
 {\em Jaynes--Cummings model} (JC)~\cite{Jaynes63} of a two-level atom strongly
coupled to a single cavity mode. Despite its success in demonstrating large single photon conditional phase
shifts, this scheme appears to be fundamentally limited by the atomic and cavity dissipation. We note, though,
the results of Hofmann \emph{et al.}~\cite{Hofmann03}, suggesting that the use of one-sided cavity can
significantly improve the phase shifts reported in~\cite{Kimble95}. Furthermore, Kojima \emph{et
al.}~\cite{Kojima03} analysed the nonlinear interaction of two photons and a two-level atom and explained
bunching and antibunching effects in the output state of photons in terms of quantum interferences between
different absorption and propagation processes.

One possible way to overcome the limitation due to dissipation is to study an {\it effective two-level system},
rather than a two-level atom. Schmidt and Imamo\u{g}lu~\cite{Schmidt96} have predicted that a four-level atomic
scheme based on electromagnetically induced transparency (EIT)~\cite{Harris97} (called EIT-Kerr scheme) can give
rise to several orders of magnitude enhancement in Kerr nonlinearity as compared to conventional two- and
three-level schemes. In this scheme, atomic spontaneous emission is avoided through EIT. The prediction has been
verified in a recent experiment by Kang and Zhu~\cite{Kang03}. It has also been predicted that the presence of
such large Kerr nonlinearities in a high-finesse cavity could result in {\em photon blockade} and effective
two-level behavior of the cavity mode~\cite{Imamoglu97,Rebic02a}. Recent progress in cavity QED demonstrates the
experimental feasibility of the observation of photon blockade using state-of-the-art cavity QED
techniques~\cite{Hood98}.

Another system predicted to exhibit photon blockade, proposed by Tian and Carmichael~\cite{Tian92}, is based on
the JC model, but involves a single two-level atom strongly coupled to the cavity mode. If the atomic and cavity
resonances coincide, and the external driving field is tuned to the lower (or upper) vacuum Rabi resonance, the
system shows characteristic two-state behaviour.

In this Brief Report, we analyse the effective two-level behaviour as exhibited by EIT-Kerr and the extended JC
schemes. By the extended JC model we mean a single two-level atom interacting with a single mode of a quantized
cavity field, where the interaction of the atom with the field mode can be described by the JC
Hamiltonian~\cite{Jaynes63}, extended by the driving and dissipation terms. The second order correlation
function $g^{(2)}(\tau)$~\cite{Walls94} has been established as a good measure of photon
blockade~\cite{Imamoglu97}, so our analysis concentrates on the properties of second order correlations. We show
that the recently developed polariton formalism~\cite{Rebic02a} can be used to account for the properties of
these correlations, provided that the model includes the entire first excitation manifold, rather than just two
levels.

\begin{figure}[t]
    \includegraphics[scale=0.65]{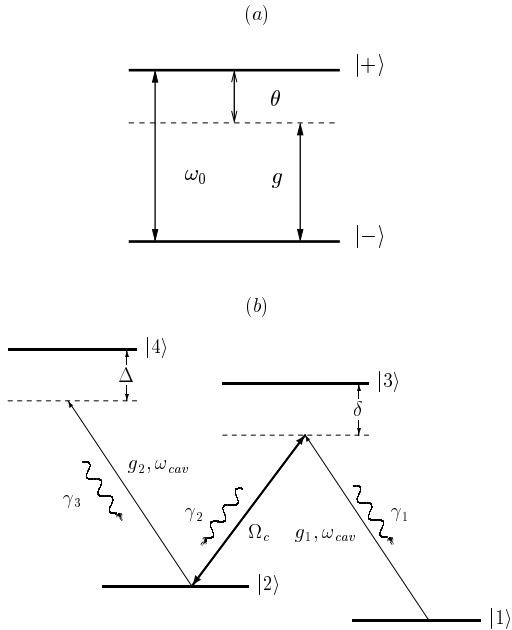}
    \caption{Energy levels in $(a)$ Jaynes-Cummings and $(b)$ EIT-Kerr
schemes. $\delta \ {\rm and} \ \Delta$ are the detunings of the field
modes from the atomic resonance, and $\gamma_j$'s are spontaneous emission
decay rates
for the given decay channel.}
    \label{fig:1}
\end{figure}

The Hamiltonian of an extended JC model in the absence of dissipation (Fig.~\ref{fig:1}$(a)$) is given by
$H_0^{JC} = H_{int}^{JC}+H_{pump}$, where
\begin{subequations}
\label{eq:jcmodel}
\begin{eqnarray}
&\! &\! \! \! H_{int}^{JC} = \hbar\theta \left(\sigma_z/2 + a^\dagger a \right)
+ i\hbar \left( g a^\dagger \sigma_- - g^*a\sigma_+ \right) \label{eq:h0jc} \\
&\! &\! \! \! H_{pump} =  i\hbar {\mathcal E}_p (a-a^{\dagger})\, .
\end{eqnarray}
\end{subequations}
Here, $H_{int}^{JC}$ is the Hamiltonian of an atom interacting with a field mode, with $\sigma$'s being the
atomic pseudospin operators, $a$ and $a^\dagger$ field annihilation and creation operators. Driving of the
cavity by a classical field of amplitude ${\mathcal E}_p$ is described by the Hamiltonian $H_{pump}$. The atom
and cavity mode (coupled with strength $g$) are assumed to be resonant, while we assume the driving field to be
detuned by $\theta = \omega_L - \omega_{cav}$ from both atomic and cavity resonance. Including dissipation leads
to the non-Hermitian Hamiltonian
\begin{equation}
  \label{eq:heffJC}
  H^{JC} = H_0^{JC} - \frac{i\hbar}{2}\kappa a^\dagger a -
  \frac{i\hbar}{2} \gamma\sigma_+\sigma_- \, ,
\end{equation}
where $\kappa {\rm \ and \ } \gamma$ are the cavity and spontaneous emission dissipation rates. Naturally,
$H^{JC}$ has to be combined with a gedanken measurement process in order to obtain the complete dynamics of the
system. This approach is usually referred to as a quantum trajectory approach~\cite{Carmichael93}.

The EIT-Kerr scheme involves a four-level atom in a cavity (see Fig.~\ref{fig:1}$(b)$). The Hamiltonian of this
model is $H_0^{EIT} = H_{int}^{EIT} + H_{pump}$, with
\begin{subequations}
\label{eq:heit}
\begin{eqnarray} H_{int}^{EIT} &=& \hbar\bigl( \delta\sigma_{33}
+\Delta\sigma_{44} \bigr) + i\hbar \bigl[ g_1\bigl(
a^{\dagger}\sigma_{13}-\sigma_{31}a \bigr) \nonumber \\
&+& \bigl( \Omega_c^*\sigma_{23}-\sigma_{32}\Omega_c \bigr)
+ g_2\bigl(a^{\dagger}\sigma_{24}-\sigma_{42}a \bigr) \bigr]
\end{eqnarray}
\end{subequations}
where $\sigma_{\mu\nu}$ are the atomic pseudospin operators, and $\Omega_c$ is the Rabi frequency of a
(classical) coupling field. Again, dissipation can be included in the same manner as above to get the
non-Hermitian Hamiltonian
\begin{eqnarray}
  \label{eq:heffEIT}
  H^{EIT} &=& H_0^{EIT} - \frac{i\hbar}{2}\bigl[ \kappa a^\dagger a
  + (\gamma_1+\gamma_2) \sigma_{33} + \gamma_3\sigma_{44} \bigr] .
\end{eqnarray}
In both cases we assume a strong atom-field coupling, leading to the
natural description in terms of dressed states (polaritons)~\cite{Rebic02a}.

The system described by the extended JC model behaves as a two-state system when excited near one of the vacuum
Rabi resonances $|e_\pm\rangle = \left(|0,+\rangle \pm |1,-\rangle\right)/\sqrt{2}$, where $-$ and + denote
ground and excited atomic states, and numbers 0 and 1 denote the number of photons in the cavity mode. The
splitting of the dressed states of the $n$'th excited manifold is found from Eq.~(\ref{eq:h0jc}), to be
$\epsilon_\pm = \frac{\hbar\theta}{2} \, n \pm \hbar g \, \sqrt{n}$. If the laser field is tuned to the lower
vacuum Rabi resonance $(\theta = g)$, the system effectively behaves as a resonantly driven two-level system.
Photon blockade occurs, since after the first photon excites the system, the second photon is detuned by
$\tilde{\epsilon} = (2-\sqrt{2}) \, \hbar g$ from the resonance of the second excitation.

It is possible to obtain the effective Hamiltonian describing the photon
blockade dynamics in terms of the two Rabi-split states $|e_\pm\rangle$.
We define the polariton operators $q_\pm$ with
$|e_\pm\rangle = q^\dagger_\pm |0,-\rangle$ and find
\begin{eqnarray}
  q_\pm \exp{(\pm i\hbar\epsilon_\pm)} = \frac{1}{\sqrt{2}}
  \left( \sigma_- \pm a \right)   \label{eq:polops} \, .
\end{eqnarray}
Substituting the operators~(\ref{eq:polops}) in the Hamiltonian~(\ref{eq:heffJC}), transforming the Hamiltonian
to a frame rotating at a laser frequency $\omega_L = \omega_{cav}-\theta$, and performing a rotating wave
approximation, we arrive at the effective Hamiltonian
\begin{eqnarray}
  \label{eq:hjceff}
  {\mathcal H}^{JC}_{eff} &=& 2\hbar\theta\, q^\dagger_+ q_+
  + i\, \frac{\hbar{\mathcal E}_p}{\sqrt{2}}\left( q_- - q_-^\dagger
  + q_+ - q_+^\dagger \right) \nonumber \\
  &\ & -i\hbar\, \frac{\gamma+\kappa}{2}\, \left( q^\dagger_-q_-
  + q^\dagger_+q_+ \right) \, .
\end{eqnarray}
In the following discussion, we assume $\theta = g$. Hamiltonian~(\ref{eq:hjceff}) contains the effective
two-level Hamiltonian of Tian and Carmichael~\cite{Tian92}, with two additional terms proportional to
$q^\dagger_+$ and $q_+$. We note three key features: $(i)$ This model is valid for large coupling $g$ and weak
driving ${\mathcal E}_p$, where the truncated (higher) manifolds do not influence the dynamics. The
applicability of the effective model can be determined by the value of $g^{(2)}(0)$, which should ideally be
zero; $(ii)$ Large amplitude oscillations in $g^{(2)}(\tau)$, of frequency $\sqrt{2}{\mathcal E}_p$, are
predicted by the effective Hamiltonian, to occur for ${\mathcal E}_p > (\kappa+\gamma)/2$; $(iii)$ Small
amplitude modulations in $g^{(2)}(\tau)$, of frequency $2\theta = 2g$ occur as a signature of the upper Rabi
resonance. The last feature points at the shortcoming of the effective two-level model (also noted by Tian and
Carmichael). Hamiltonian~(\ref{eq:hjceff}) represents an {\em effective two-manifold model}, reducing the
dynamics to transitions between the ground state and the entire first excited manifold.

Dressed states analysis for a single atom in the EIT-Kerr configuration has been carried out in
Refs.~\cite{Imamoglu97,Rebic02a}. There are three states in the $n=1$ manifold, one of which is resonant with
the cavity mode. The second manifold contains four states. The outer two states are detuned far from the
resonance and therefore their contribution to the system dynamics is negligible. The inner two states are also
detuned, but lie closer to the resonance, with the size of the detuning determined by a coupling strength $g_2$.

The three states in the first excited manifold are
\begin{subequations}
  \label{eq:estates}
\begin{eqnarray}
  |\phi_0\rangle &=& \frac{|1,1\rangle +
  (g_1/\Omega_c)|0,2\rangle}{\sqrt{1+(g_1/\Omega_c)^2}} \label{eq:estates1} \\
  |\phi_\pm\rangle &=& -\frac{(g_1/\Omega_c)|1,1\rangle
  + i(\varepsilon_\pm/\Omega_c)|0,3\rangle
  - |0,2\rangle}{\sqrt{1+(\varepsilon_\pm/\Omega_c)^2+(g_1/\Omega_c)^2}}
  \label{eq:estatespm}
\end{eqnarray}
\end{subequations}
Again, the effective two-manifold model can be obtained by following the
same method. Defining the polariton operators $p_j$ with
$|\phi_j\rangle = p_j^\dagger |0,1\rangle$, the following effective
Hamiltonian emerges
\begin{eqnarray}
    \label{eq:heiteff}
    {\mathcal H}_{eff}^{EIT} &=& \hbar\varepsilon_-\, p_-^\dagger p_-
    + \hbar\varepsilon_+\, p_+^\dagger p_+ + i\frac{\hbar\Omega_-}{2}
    \bigl( p_- - p_-^\dagger \bigr)    \nonumber \\
    &\ &+ i\frac{\hbar\Omega_+}{2} \bigl( p_+ - p_+^\dagger \bigr )
    + i\frac{\hbar\Omega_R}{2} \bigl( p_0 - p_0^\dagger \bigr)   \nonumber \\
    &\ &+i\hbar\Gamma_0 p_0^\dagger p_0 + i\hbar\Gamma^{(1)}_- \,
    p_-^\dagger p_- + i\hbar\Gamma^{(1)}_+ \, p_+^\dagger p_+ \, ,
\end{eqnarray}
with the effective Rabi frequencies of driving~\cite{Rebic02a},
\begin{subequations}
  \begin{eqnarray}
    \label{eq:factors}
    \Omega_\pm &=& -2{\mathcal E}_p\,
    \frac{g_1/\Omega_c}{\sqrt{1+(\varepsilon_\pm/\hbar\Omega_c)^2+(g_1/\Omega_c)^2}}
    \label{eq:omegapm} \, , \\
    \Omega_R &=& \frac{2{\mathcal E}_p}{\sqrt{1+\bigl( g_1/\Omega_c\bigr)^2}}
    \label{eq:omegar} \, ,
  \end{eqnarray}
decay rates,
  \begin{eqnarray}
    \Gamma_0 &=& \frac{\kappa}{1+\left(g_1/\Omega_c \right)^2}
    \label{eq:gamma0} \, , \\
    \Gamma^{(1)}_\pm &=& \frac{\kappa g_1^2
    + (\gamma_1+\gamma_2)\left(\varepsilon_\pm \right)^2}
    {g_1^2+\Omega_c^2+\left(\varepsilon_\pm \right)^2}\, .
    \end{eqnarray}
\end{subequations}
and energies $\varepsilon_\pm = \delta/2 \pm \sqrt{\left(\delta/2 \right)^2 + \Omega_c^2 + g_1^2}$. Key features
of this effective two-manifold model can be identified in correspondence to those of the extended JC model.
Large oscillations of frequency $\Omega_R$ in $g^{(2)}(\tau)$ are predicted to occur for $\Omega_R >
\Gamma_0/2$, or ${\mathcal E}_p > (\kappa/4)/\sqrt{1+g_1^2/\Omega_c^2}$. In addition, small amplitude
modulations consisting of {\em two} frequencies $\varepsilon_\pm$ will be observed. If $|\varepsilon_+| \approx
|\varepsilon_-|$, only one frequency will be visible, whereas if $|\varepsilon_+| \neq |\varepsilon_-|$,
oscillations with both frequencies should be apparent.

\begin{figure}[t!]
    \hspace{-5mm}
    \includegraphics[width=\linewidth,height=\linewidth]{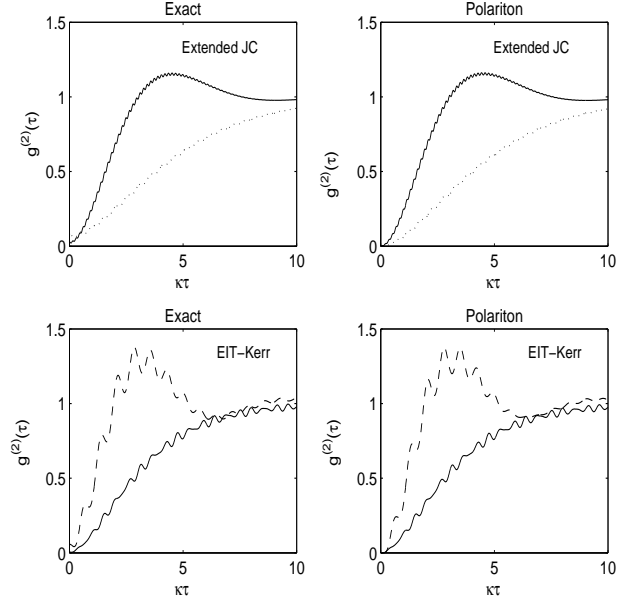}
    \caption{Second order correlation functions in Jaynes-Cummings and EIT-Kerr
schemes. Left column (``Exact'') is obtained from the exact models, while right column (``Polariton'') from the
effective two-manifold models. In the ``Exact'' case, parameters for the extended JC are $\gamma = 0.1\kappa$,
$\theta = g$ and $g=6\kappa$, ${\mathcal E}_p = 0.1\kappa$ for the dashed curve; $g=20\kappa$, ${\mathcal E}_p =
0.5\kappa$ for the solid curve. For the EIT-Kerr scheme, parameters are $\gamma_j = 0.1\kappa$, $g_j=6\kappa$,
$\Delta = 0$ and $\delta = 0.2\kappa$, ${\mathcal E}_p = 0.7\kappa, \ \Omega_c = 6 \kappa$ (dashed); $\delta =
4\kappa$, ${\mathcal E}_p = 0.1\kappa$, $\Omega_c = 12\kappa$ (solid). The same parameters are used in the
``Polariton'' case. In the EIT-Kerr case, this amounts to $\varepsilon_+ = 8.59\kappa$, $\varepsilon_- =
-8.39\kappa$, $\Omega_+ = -0.696\kappa$, $\Omega_- = -0.704\kappa$, $\Omega_R = 0.99\kappa$, $\Gamma_+ =
0.348\kappa$, $\Gamma_- = 0.352\kappa$ and $\Gamma_0 = 0.5\kappa$ (dashed); $\varepsilon_+ = 15.56\kappa$,
$\varepsilon_- = -11.56\kappa$, $\Omega_+ = -0.06\kappa$, $\Omega_- = -0.06\kappa$, $\Omega_R = 0.18\kappa$,
$\Gamma_+ = 0.2\kappa$, $\Gamma_- = 0.2\kappa$ and $\Gamma_0 = 0.8\kappa$ (solid).}
    \label{fig:2}
\end{figure}
How well do these effective Hamiltonians describe the dynamics of the full system? We calculate the second order
correlation function using a wave function simulations~\cite{Carmichael93} of the original
Hamiltonians~(\ref{eq:heffJC}) and~(\ref{eq:heffEIT}). The photon space was restricted to 4 photons, resulting
in a 10 (20) dimensional Hilbert spaces for the extended JC (EIT-Kerr) model.  Results are depicted in
Fig.~\ref{fig:2}. Then, using the same technique, we calculate the second order correlation function of the
effective polariton Hamiltonians~(\ref{eq:hjceff}) and~(\ref{eq:heiteff}), requiring Hilbert spaces of only 3
and 4 dimensions respectively, and also plot the results in Fig.~\ref{fig:2}.

Identical values of couplings and decay rates are chosen in both schemes to
enable better comparison. In the extended JC scheme, a significant antibunching
(as measured by $g^{(2)}(\tau = 0)$) is found. The particular value of $g^{(2)}(0)$
measures the validity of the truncation of dressed basis after the first manifold.
To achieve even better agreement, a stronger coupling is needed
($g/\kappa \sim 20$ or larger) which is experimentally unavailible as of yet.
We note the modulation of frequency $2g$, as predicted by two-manifold model.

The other two curves show the simulation results for the EIT-Kerr system. They exhibit essentially the same
values at the origin, which are now very close to zero. This means that the effective
Hamiltonian~(\ref{eq:heiteff}) captures the significant dynamics well. The modulation of frequency
$\varepsilon_+ \approx |\varepsilon_-| \approx \sqrt{g_1^2+\Omega_c^2}$, as predicted by the effective
Hamiltonian is seen on the upper curve. The driving chosen to produce Fig. \ref{fig:2} is such that $\Omega_R >
\Gamma_0$, implying the presence of large oscillations in the correlation function. The opposite is true for the
lower curve. Note also the existence of two modulation frequencies since the choice $\delta = 4\kappa$ implies
$\varepsilon_+ \neq |\varepsilon_-|$.

The two curves in the EIT-Kerr case show vastly different coherence times
-- a difference
attributable to the lifetime of the effective excited state~(\ref{eq:estates1}).
It follows from the decay rate~(\ref{eq:gamma0}) that, given the fixed atom-field
coupling $g_1$, the lifetime can be adjusted by the coupling laser (i.e., $\Omega_c$).
As a consequence, the coherence time of the effective two-level system can be
adjusted to virtually any prescribed value by varying the coupling $\Omega_c$.
Indeed, for larger values of $g_1/\Omega_c$, such as the one depicted in the lower
EIT-Kerr curve in Fig.~\ref{fig:2}, there is a comparatively slow recovery of the
correlation function from the origin.

In conclusion, we have presented a polariton description of effective two-level
atom-cavity systems in the strong coupling regime of cavity QED.
It was shown how a
reduction of a more sophisticated polariton structure to a lowest excited manifold
can account for the properties of a second order correlation function of light
leaking out of the cavity.

\begin{acknowledgments}
The authors would like to thank A. Imamo\u{g}lu for stimulating discussions and
the Marsden Fund of the Royal Society of New Zealand for the financial support.
\end{acknowledgments}

\end{document}